\documentclass[aps,prl,twocolumn,groupedaddress]{revtex4}

\usepackage{graphics} 
\usepackage{graphicx}
\usepackage{dcolumn}
\usepackage{bm}
\usepackage{epstopdf}
\usepackage{amsmath}
\usepackage{bm}
\everymath{\displaystyle}
\usepackage{xcolor, soul}
\sethlcolor{yellow}

\begin{document}

\title{ENTROPY-BASED THEORY OF THERMOMAGNETIC PHENOMENA: \\
Poynting Vector, Vorticity, and Advanced Sensing}

\author{Andrei Sergeev }
\email[Corresponding author: ]{podolsk37@gmail.com}
\affiliation{*U.S. Army Research Laboratory, Adelphi, Maryland 20783, USA}
\author{Michael Reizer}
\affiliation{Baruch Hashem Beseder Lab}
\email[Corresponding author: ]{mreyzer@gmail.com}
\date{\today}

\begin{abstract}
We show that in the linear response approximation only entropy provides coupling between thermal and electric phenomena. The dissipationless quantum currents, - magnetization, superconducting, persistent and topological edge currents, - do not produce and transfer entropy and may be excluded from final formulas for thermomagnetic coefficients. The magnetization energy flux, $c \bf{M} \times \bf{E}$, in crossed electric and magnetic fields strongly modifies the Poynting vector in magnetic materials and metamaterials, but do not contribute to the heat current. Calculating entropy fluxes of fluctuating Cooper pairs, we find the fluctuation Nernst coefficient in pure superconductors. To account electron scattering, we generalize the gauge-invariant  Kubo formalism developed for the Hall effect to thermomagnetic entropy transfer.  We also introduce the thermomagnetic entropy per unit charge and derive the Nernst coefficient proportional to the difference of the thermoelectric and thermomagnetic entropies. This explains the Sondheimer cancellation and high sensitivity of thermomagnetic phenomena to correlations. In 2D superconductors, the transport entropy transferred by a vortex moving through the background formed by vortex-antivortex pairs is the configuration entropy of $k_B \cdot \ln 2$, which strongly exceeds the intrinsic entropy of vortex core. Beyond the linear response, the non-entropic forces can lead to phenomena unexpected from thermodynamics, such as vortex attraction to the moving hot spot. Quantum currents do not transfer entropy and may be used as ideal connectors to quantum nanodetectors.
\end{abstract}
\maketitle

\section{Introduction}

In 1886, Albert von Ettingshausen and Walther Nernst while studying the Hall effect in bismuth also investigated the ``thermal Hall effect.'' Instead of creating electric current in the bismuth ribbon placed in the magnetic field, they warmed one end of the ribbon and generated the heat flow from the hot to the cold end of the ribbon.\cite{1} The discovered phenomenon, - induction of electric current or voltage in the direction perpendicular to both magnetic field and the heat current, - is currently known as the Nernst effect. These experimental investigations were carried out under the supervision of Boltzmann, who realized the entropic nature of thermal forces and foreseen the related phenomenon, - generation of the transverse entropy flow (heat current) or temperature gradient due to magnetic field and the electric current. In 1931, considering entropy production, Onsager derived reciprocal relations for certain coefficients between flows and forces, including entropy flow and entropic forces generated by the temperature gradient.\cite{2} In 1948 Callen applied the Onsager's relations to thermomagnetic phenomena and derived exact formal formula for thermomagnetic coefficients.\cite{3} 
  
Thermomagnetic phenomena observed in wide variety of conducting materials, including metals, semiconductors, organic materials, and superconductors.\cite{4,5,6, 7} In most metals and semiconductors, thermomagnetic effects are well described by the model of noninteracting electrons.\cite{4,8}  In this model, formulas the thermomagnetic transport coefficient in metals may be obtained by replacing the electron density in the formulas for the Hall conductivity by the electron entropy. For semiconductors, electron and hole densities should be replaced by the electron and hole entropies.  For a degenerate conductor, the electron entropy brings the degeneracy factor of $k_B T/\epsilon_F$, where $\epsilon_F$ is the Fermi energy.  Therefore, the Fermi statistics puts strong limitations on thermomagnetic coefficient and significant thermomagnetic effects are expected in conductors with small value of the Fermi energy.\cite{9} 

Novel strongly correlated materials turn out to be a great challenge for the theory of thermomagnetic phenomena. 
 Giant thermomagnetic effects observed in CeCoIn$_5$, Bechgaard salts, URu$_2$Si$_2$ and PrFe$_4$P$_{12}$  are usually linked with exotic electronic orders and interaction effects are described in terms specific configuration entropy transferred by electrons.\cite{10,11,12}  Problems of microscopic theory of entropy flux for interacting electrons were discussed in a recent review by Kadanoff.\cite{13} In any approximation for electron correlations, the transport entropy is defined via combination of the distribution functions (Green's functions), which are affected by correlations.  While this approach does not work for the entanglement entropy, due to its strongly nonlocal nature, it is applicable to many popular physical models.\cite{13} Anyway, the correlation energy by itself cannot be equated to the heat. 
 
Meanwhile numerous recent papers directly associate giant thermomagnetic effects with the flow of large correlation energy, which has quantum and/or electromagnetic origin. In particular, many papers\cite{14,15,16,17,18,19,20,21} associate the flux of magnetization energy, $c \bf{M} \times \bf{E}$,  in crossed magnetic and electric fields with the heat (entropy) flow. In the electron systems with large magnetization currents, e.g. near the superconducting transition, the obtained magnetization contribution to the thermomagnetic coefficient may exceed the thermomagnetic coefficient in metals  by the factor of $\epsilon_F/k_B T$.\cite{14,15,16,17,18,19,20,21,22} The corresponding thermal forces are associated with the electromagnetic interaction between an electron and the magnetization currents generated by the temperature gradient, $-c(d \bf{M} /dT) \times \nabla T$.  At the same time, it is well understood that the coherent electron motion in magnetization, superconducting, persistent and topological edge currents has quantum origin and the entropy of any coherent current is zero.\cite{23,24,25,26,27,28}  The nonzero transport entropy of magnetization and superconducting currents is a long term enigma. 
 
In particular, above the superconducting transition, the fluctuating superconducting magnetization currents are large and the contribution of $c \bf{M} \times \bf{E}$ to the thermomagnetic heat flow is huge. To satisfy the Onsager relations modified by magnetization contribution, many theoretical papers obtained the ''giant'' Nernst coefficient due to  the Gaussian fluctuations of Cooper pairs.\cite{14,15,16,17,18,19,20,21,29,30,31,32,33} In a weak magnetic field the fluctuation contribution to the Nernst coefficient exceeds its value in a normal state by a huge factor of  $(\epsilon_F/ k_B T)^2$. At the same time, it is well understood that superconducting fluctuations give small corrections to the electron entropy. Therefore, the origin of this giant fluctuation Nernst effect is still unclear. Some papers\cite{34,35,36} associate it with the strong temperature-dependence of the thermodynamic chemical potential of Cooper pairs.  However, the thermodynamic chemical potential of Cooper pairs is always zero due to the lack of constraint on the number of Cooper pairs in the system.\cite{23} For the same reason, the chemical potential of the photons in the equilibrium black-body radiation is zero, while a non-zero chemical potential  is reached in solar cells.\cite{37,38} Thermodynamic limitations on quasi-equilibrium (the linear response) and strongly nonequilibrium thermomagnetic transport are not well understood yet. 

The electromagnetic energy and thermal energy balance equations could shed some light on the form of the heat current in the magnetic field. However, this question also turns out to be rather complicated. Recent theories associate the magnetization currents in electric field with dissipation and entropy production.\cite{15,39,40} Assumption of the dissipative character of magnetization currents immediately leads to the magnetization contribution to the Poynting vector of $c \bf{M} \times \bf{E}$  and, therefore, the Poynting vector in magnetic materials has a form $c [{\bf E} \times {\bf B}]/4 \pi$.\cite{39,40} While this modified form of the Poynting vector is widely accepted in the area of metamaterials, some authors\cite{41} find that this result contradicts  optical properties of metamaterials and this question is still far from resolution. 

In this work, we develop consistent entropy-based description of thermomagnetic phenomena and investigate a role of magnetization and superconducting currents in thermomagnetic transport and energy transfer (Poynting vector).  
We show that entropy and only entropy provides coupling between thermal and electric phenomena in the linear response approximation. Non-entropic energy fluxes and non-entropic thermal forces do not contribute to thermomagnetic phenomena. In Section II, we consider the textbook model of noninteracting electrons in quantized magnetic field and show that the magnetization energy flux, $\bf{M} \times \bf{E}$, does not transfer the entropy. In Section III, we discuss entropic and non-entropic thermal forces induced by the temperature gradient and show that  to satisfy the Onsager relations the non-entropic forces should be canceled among themselves. In Section IV we review a divergent-free nature of magnetization currents, which leads to cancellation of bulk and surface magnetization currents\cite{42} and cancellation of corresponding non-entropic forces in thermomagnetic phenomena.  In Section V we consider transfer of magnetization energy, $\bf{M} \times \bf{E}$,  and derive a new formula for the Poynting vector in magnetic materials and metamaterials. In Sections VI and VII we investigate fluctuation thermomagnetic phenomena in superconductors above the transition. In the collisionless limit, we derive exact formulas  for the thermomagnetic coefficient, which is proportional to electron entropy corrected due to superconducting fluctuations. To calculate the fluctuation Nernst coefficient in a disordered superconductor, we employ the gauge - invariant Kubo formalism developed for the fluctuation Hall effect.\cite{43} 

In Section VIII we generalize the entropic  approach to electrons with finite mobility and derive exact thermodynamic equations, which correct recently proposed phenomenological  formulas for the Nernst coefficient.\cite{34,35,36}  In Section IX we consider thermomagnetic transport in the vortex liquid.  While, in accordance with the Onsager approach, non-entropic thermal forces are canceled  among themselves in quasi-equilibrium conditions, this cancellation is not valid for the local or time modulated temperature gradient. In this context we review and explain recent experiments that observe motion of single superconducting vortices, which follow to the hot spot generated by a laser beam.\cite{44}  We also discuss thermomagnetic phenomena in the Berezinskii - Kosterlitz - Thouless (BKT) state of 2D superconductors and show that the transport entropy transfered by a vortex moving through the background formed by vortex-antivortex pairs is the configuration entropy of $k_B \cdot \ln 2$. Finally,  the unique property of quantum currents that transfer electric charge without entropy is discussed in Section X for applications to ultra-sensitive detectors. 

\section{Entropy and Magnetization Fluxes in Collisionless Thermomagnetic Transport}

Thermomagnetic coefficient, $\mathcal{L}$, describes the transport electric current as a response to the temperature gradient in the magnetic field, 
\begin{eqnarray}
{\bf j}^e_{tr} = \mathcal{L} \cdot { [ \nabla T \times  {\bf H}]  \over H}. 
\label{1} 
\end{eqnarray}
Taking into account the Onsager relation, the corresponding equation for the heat current  in crossed electric and magnetic fields has a form, 
\begin{eqnarray}
{\bf j}^h = T\mathcal{L} \cdot { [{\bf H} \times {\bf E}] \over H } 
\label{2} 
\end{eqnarray}
The Nernst effect may be considered as the transverse thermopower in the magnetic field. The Nernst coefficient,  $\mathcal{N}$, relates the electric field induced in the open circuit by crossed temperature gradient and magnetic field, 
\begin{eqnarray}
 \mathcal{N} = { \nabla \varphi_\bot  \over H \cdot \nabla T }   = {1\over H} {  ( \mathcal{L}  + \mathcal{S} \sigma_H) \sigma \over \sigma^2 +\sigma_H^2}, 
\label{3}  
\end{eqnarray}
where $ \varphi_\bot$ is the measured voltage, $\sigma$ is the electrical conductivity, $\sigma_H$ is the Hall conductivity, and $\mathcal{S}$ is the thermopower or Seebeck coefficient. In particular, in the Fermi liquid with the energy independent electron scattering, the transport coefficients are related as 
$\mathcal{L} = \mathcal{S} \sigma_H $ and, as a result, the Nernst coefficient is zero. This fact is well known as Sondheimer cancellation.\cite{4} In general, for the Fermi liquid in a weak magnetic field both the thermomagnetic coefficient $\mathcal{L}$ and the product of the Seebeck and Hall coefficients are proportional to product of $\omega_c \tau$ and $k_BT/\epsilon_F$, and,  therefore, the Nernst coefficient is  very small. 

The Ettingshausen effect may be considered as the thermal Hall effect and the Ettingshausen coefficient is defined as 
\begin{eqnarray}
 \mathcal{E} = { \nabla T_\bot \over  H \cdot { j}^e_{tr}  } . 
\label{4} 
\end{eqnarray}
Due to the Onsager relation given by Eqs. 1 and 2, the Nernst and Ettingshausen coefficients are interconnected by a fundamental relation obtained by Bridgman,\cite{45} 
\begin{eqnarray}
 \mathcal{E} = { T \over \kappa}  \mathcal{N}, 
\label{5} 
\end{eqnarray}
where $\kappa$ is the thermal conductivity. Thus, all measurable transverse thermomagnetic effects may be presented via the thermomagnetic coefficient, $ \mathcal{L}$, and other well-known coefficients that describe the electric and thermal transport in the absence of magnetic field. 

Numerous recent theoretical papers state that magnetization current directly contribute to the transverse heat current induced by electric field and in some materials the magnetization heat current,  can provide significant contribution to the thermomagnetic effects. According to these works\cite{14,15,16,17,18,19,20,21} the heat current consists of the transport and magnetization components and the magnetization heat current is given by  
\begin{eqnarray}
{\bf J}^h &=& {\bf J}^h_{tr} + {\bf J}^h_{mag} \\
 {\bf J}^h_{mag} &=& - c[{\bf E}\times {\bf M}(T) ],
\label{6,7} 
\end{eqnarray}
where ${\bf M}$ is the magnetization and ${\bf E}$ is the electric field.  Eqs. 6 an 7 were empirically proposed in Ref. 14 to satisfy the Onsager relation for heat and electric bulk thermomagnetic currents above the superconducting transition.  The same equations  were derived in the book\cite{15} on a base of assumption that magnetization currents in the electric field dissipate the electromagnetic energy in the same way as the transport currents, i.e. the dissipated power is given by ${\bf j}_{tr} \cdot {\bf E} + {\bf j}_{mag} \cdot {\bf E}$.

Here we will show that Eqs. 6 and 7 are wrong. The energy flux
\begin{equation}
{\bf P}_{mag} =-c [{\bf E} \times {\bf M}(T) ],
\label{8} 
\end{equation} 
has nothing to do with the heat current. In Section V we will show that  ${\bf P}_{mag}$ is the magnetization part of the Poynting vector, which describes the transfer of electromagnetic energy by the bulk magnetization currents. 

First of all, magnetization currents have zero intrinsic entropy. The magnetization currents are generated by magnetic vector potential without the electrical electromotive force. Therefore, the magnetization currents have a quantum nature. In other words, in accordance with the Bohr - van Leeuwen theorem,\cite{46} the magnetization currents would vanish, if the motion of electrons is considered within the classical mechanics. Magnetization current are not much sensitive to scattering of electrons, because the scattering only slightly shifts the electron orbits without substantial changes of the phase of the electron wave function. In terms of thermodynamics, the entropy related to the electromagnetic magnetization part of the free energy, $F_{em}$ is given by\cite{24}
\begin{equation}
 S_{mag} = \left(  {\partial F_{em} ( M(T) )\over \partial T} \right)_M = 0,
\label{9} 
\end{equation}
because the temperature derivative is calculated at constant magnetization and the temperature dependence of ${\bf M}$ does not contribute to the entropy.\cite{26} 

The direct relation between the entropy and thermomagnetic coefficients may be obtained in the collisionless limit. If the electron mean free path, $\ell$, exceeds the Larmor radius, $r_L$, one can neglect electron scattering.  In this case the whole electron system moves with the drift velocity, 
\begin{equation}
{\bf v}_{dr} =  c \cdot { [{\bf F} \times {\bf H}] \over e H^2},
\label{10} 
\end{equation} 
where ${\bf F}$ is the force, which is perpendicular to the magnetic field, ${\bf H}$, $c$ is the light velocity, and $e$ is the electron charge. In particular, in crossed electric and magnetic fields, the electron drift velocity is  $c [{\bf E} \times {\bf H}] / H^2$. Taking into account that the electron thermal energy is $k_B T S_e $ ($S_e$ is the electron entropy), the heat current may be presented as
\begin{equation}
{\bf J}^{h}_{tr} = {\bf v}_{dr} TS_e = c  TS_e \cdot { [{\bf E} \times {\bf H}] \over H^2},
\label{11} 
\end{equation} 
and the thermomagnetic coefficient is given by 
\begin{equation}
  \mathcal{L} = - c \cdot {S_e \over H}. 
\label{12} 
\end{equation} 
Thus, in the collisionless limit the thermomagnetic coefficient is directly expressed via the thermodynamic entropy of electron subsytem. Eq. 12 is well known in the theory of thermomagnetic phenomena.\cite{8}  

Let us apply this equation to non-interacting electrons in quantized magnetic field and  calculate the oscillating part of the thermomagnetic transport coefficient. The entropy may be calculated from the thermodynamic potential, which is given by\cite{47}
\begin{eqnarray}
&&\Omega (T, H, \zeta) = {m^{3/2} (\hbar \omega_c)^{5/2} \over 4 \pi^4 \hbar^3   } \\ \nonumber &&\times \sum_{k}
(-1)^k k^{-5/2} \cdot \Psi(\lambda\cdot k) 
\cdot \cos\left( {2\pi \zeta \over \hbar \omega_c}\cdot k - {\pi\over 4}  \right),
\label{13} 
\end{eqnarray}
where $\omega_c =eH/mc$ is the cyclotron frequency, $\lambda(T,H) = 2 \pi^2 k_B T/\hbar \omega_c$, 
$\Psi(x) = x/\sinh(x)$, and $\zeta$ is the electron chemical potential. In Eq. 13 we limit our consideration by the oscillating part of the thermodynamic potential.
Calculating the entropy, 
\begin{equation}
 S = \left(  {\partial \Omega(T,H, \zeta) \over \partial T} \right)_{H,\zeta} \   \ ,
\label{14} 
\end{equation} 
and substituting it in Eg. 12, we get the thermomagnetic  coefficient of free electrons in quantized magnetic field,
\begin{eqnarray}
 &&\mathcal{L} = {ek_B m^{1/2} (\hbar  \omega_c )^{1/2} \over 2 \pi^2 \hbar^2   } \\ \nonumber &&\times  \sum_{k}
(-1)^k k^{-3/2} \cdot \Psi' (\lambda\cdot k) 
\cdot \cos\left( {2\pi \zeta \over \hbar \omega_c}\cdot k - {\pi\over 4}  \right).
\label{15} 
\end{eqnarray}
Eq. 15 coincides with the result obtained by Bar'yachtar and Peletminskii using the linear response approach.\cite{48}

Now we compare the thermomagnetic heat current (Eq. 15) with the electromagnetic flux related the magnetization. In the limit of quantized field, the magnetization is given by\cite{47}
 \begin{eqnarray}
&& M = \left(  {\partial \Omega(T,H, \zeta) \over \partial  H} \right)_{T,\zeta} 
 = -  {e m^{1/2} (\hbar  \omega_c )^{1/2} \zeta \over 2 \pi^3 \hbar^2  c } \\ \nonumber
&& \times \sum_{k}
(-1)^k k^{-3/2} \cdot \Psi(\lambda\cdot k) 
\cdot \sin\left( {2\pi \zeta \over \hbar \omega_c}\cdot k - {\pi\over 4}  \right). 
\label{16} 
 \end{eqnarray}  
Comparing the thermomagnetic heat current, ${\bf j}^h = T\mathcal{L} \cdot [{\bf H} \times {\bf E}] /H$, where $\mathcal{L}$ is given by Eq. 15, and  the magnetization part of the Poynting vector ${\bf j}^\epsilon_{mag} = c[{\bf M}\times {\bf E} ]$, where $M$ is given by Eq. 16, we see that the electromagnetic flux exceeds the heat flux by the factor of $\zeta / k_B T$.  If we would follow to Refs. \cite{14,15,16,17,18,19,20,21} and associate the electromagnetic flux with the heat, we obtain another ''giant''  Nernst effect without the degeneracy factor of $k_B T/\epsilon_F$ for free electrons in quantized magnetic field.
Obviously, this conclusion is erroneous. 
\par\smallskip 
Thus, in the collisionless limit, the thermomagnetic coefficient may be directly presented via the entropy of electrons (Eq. 12).   A textbook example of electrons in quantized magnetic field shows that the transverse flux of magnetization energy, $c[{\bf M} \times {\bf E}]$, drastically exceeds the transverse heat current and, therefore, the magnetization energy cannot be associated with the heat. 

\section{Thermomagnetic Charge Transfer Induced by Entropic Forces}

In this section we discus a thermal force induced by the temperature gradient and the Onsager relation for thermomagnetic coefficients. To calculate the thermal force, let us divide the electron system into a large number of small statistical subsytems of electrons. So, the total number of electrons is given by $N_e = \Sigma_i \Delta N_i $ and the total entropy is $S_e = \Sigma_i \Delta S_i $. When one of this subsystems moves from the point ${\bf R}$ with temperature $T$ to the point ${\bf R} + \Delta{\bf R}$ with temperature $T +\Delta T$, the entropic force produces the work of $A = {\bf F}^{}_{ i} \cdot \Delta {\bf R} $ and transfers the thermal energy of $Q = T\cdot \Delta S_i$  . According to the second law of thermodynamics, 
\begin{eqnarray}
{A \over Q} = {{\bf F}^{th}_{ i} \cdot \Delta {\bf R} \over T\cdot \Delta S_i} = - {\Delta T \over T}. 
\label{17} 
\end{eqnarray}
From the last equation we obtained that the thermal force acting on the subsystem is ${\bf F}^{th}_{ i} =- \Delta S_i \cdot \nabla T  $, and the thermal force per electron in this subsystem is 
\begin{eqnarray}
{\bf f}^{th}_{ i} =- {\Delta S_i \over \Delta N_i} \cdot \nabla T. 
\label{18} 
\end{eqnarray}
As in the previous Section, we limit our consideration by the colisionless limit. In this case the drift velocity of electrons is given by 
\begin{eqnarray}
{\bf v}_{dr} =- {c\over e} {[{\bf f}^{th}_{ i}\times {\bf H}] \over H^2}  = - {c\over e}  {\Delta S_i \over \Delta N_i} \cdot 
{ \nabla T \times {\bf H} \over H^2}. 
\label{19} 
\end{eqnarray}
 Thus, the electric current transferred by the subsystem is    
\begin{eqnarray}
\Delta {\bf j}^e_i  =   e {\bf v}_{dr} \cdot \Delta N_i = - c \Delta S_i \cdot  { \nabla T \times {\bf H} \over H^2},
\label{20} 
\end{eqnarray}
and the total thermomagnetic transport electric current is  
\begin{eqnarray}
{\bf j}^e_{tr} = - c S_e \cdot  { \nabla T \times {\bf H} \over H^2},
\label{21} 
\end{eqnarray}
Eqs. 12 and 21 directly demonstrate validity of the Onsager relations in thermomagnetic phenomena (Eqs. 1 and 2). If the temperature gradient is taken into account via thermal entropic force (Eq. 18), the obtained electric current does not have any magnetization component (Eq. 21), because magnetization currents have zero intrinsic entropy. In other words, the electric transport current is directly calculated as a response to thermal force. This approach was proposed by  Stephen\cite{49}  for description of  thermomagnetic phenomena in the superconducting vortex state and it is widely used  in this area.\cite{26,50}

To the best of our knowledge, the entropic approach based on thermal forces was not used in microscopic theories. Therefore, microscopically calculated electric current induced by the temperature gradient incorporates contributions from the all temperature dependent material characteristics, such as magnetization.  
 In particular, in the Boltzmann transport equation the temperature gradient is usually introduced via the dependence of $T(x)$ in the argument, $\varsigma = (\epsilon ( {\bf p}) - \mu) / T(x) $ of the electron distribution function $f(\varsigma)$. In the collisionless limit, in the first order in $\nabla T$ the nonequilibrium electron distribution function has well-known form,\cite{8} 
\begin{eqnarray}
f_1 ({\bf p})  ={\epsilon - \zeta \over T} \cdot {\partial f_0 \over \partial \epsilon} \cdot  {v_y^2 \over \omega_c} \cdot \nabla_x T,  
\label{22} 
\end{eqnarray}
where $v_y$ is the $y$-component of the electron velocity, and $\omega_c$ is the cyclotron frequency. 

In the quantum transport equation, besides the nonequilibrium distribution function proportional to the temperature gradient, one should include quantum corrections in the form of the Poisson brackets calculated with any two electron Green functions  (polarization operators, fluctuations propagators etc), $A(\omega, {\bf q}, T)$ and $B(\omega, {\bf q}, T)$, that correspond to different time intervals in the diagrammatic technique,\cite{27,51,52,53}
 \begin{eqnarray}
 \{A(\omega, {\bf q}, T), B(\omega, {\bf q}, T)  \} =  {i \over 2} \nabla T \left( {\partial A\over \partial T} {\partial B \over \partial {\bf p}}  - {\partial B \over \partial T}  {\partial A\over \partial {\bf p}}  \right). 
\label{23} 
\end{eqnarray}

The $\nabla T$-induced electric current calculated in the Boltzmann formalism or in the quantum transport approach will include
contributions of all temperature dependent characteristics, such as magnetization, Cooper pair density etc.  In the case of the temperature dependent magnetization, ${\bf M}(T)$,
 the electric bulk current is given by      
\begin{eqnarray}
{\bf j}^e_{bulk} = {\bf j}^e_{tr} +{\bf  j}^e_{mag}, 
\label{24} 
\end{eqnarray}
where the magnetization current density in the homogeneous sample is 
\begin{eqnarray}
{\bf j}^e_{mag} =  c \nabla \times {\bf M}(T) = c \nabla T  \times  {\partial {\bf M}(T) \over \partial T}.
\label{25} 
\end{eqnarray}
In other words, the thermomagnetic transport current is given by 
\begin{eqnarray}
{\bf j}^e_{tr} = {\bf j}^e_{bulk} + c  {\partial {\bf M}(T) \over \partial T}  \times  \nabla T.
\label{26} 
\end{eqnarray}
Eq. 26 was derived by Obraztsov,\cite{54} who has pointed out that to satisfy the Onsager relation presented by Eqs. 1 and 2
the bulk electric current generated by the temperature gradient should be corrected by magnetization currents. Employing  the quantum transport equation, in our work\cite{27} we have confirmed that bulk thermomagnetic electric current should be corrected by magnetization currents, while the heat bulk current does not have such corrections. 

As we have shown in the previous section, the magnetization currents do not contribute the thermomagnetic heat transfer. In the first part of this section, we have demonstrated that the magnetization currents do not appear in the thermomagnetic charge transfer, which is produced by the entropy-related thermal force  generated by $\nabla T$ (Eq. 18). Therefore, magnetization currents do not arise in the thermomagnetic theory formulated in terms of entropy and thermal forces. If  $\nabla T$ is taken into account in the electron distribution function (Green function Poisson brackets in quantum formalism), the temperature-dependent magnetization currents, ${\bf j}^e_{mag} (T) $,  give contribution to the bulk electric current. To obtained electric transport current, ${\bf j}^e_{tr}$, the magnetization contribution should be removed from the bulk current (Eq. 26). 

The textbook example of electrons in quantized magnetic field considered in the previous section shows that the magnetization correction to the electric current, $   c\cdot ({\partial  M/ \partial T)}  \times  \nabla T$, where $M$ is given by Eq. 16, drastically exceeds the thermomagnetic transport current. Therefore, in the Boltzmann type formalism it is important accurately calculate the magnetization component in the bulk current and then remove it, in accordance with the Obraztsov's formula (Eq. 26), to find the transport current.  If one would neglect the thermomagnetic component in the bulk current of free electrons in quantized magnetic field, the obtained thermomagnetic coefficient $  \mathcal{L} = c\cdot (\partial  M / \partial T)$ exceeds the correct coefficient given by Eq. 15 by a huge factor of $\epsilon_F /T$.  Such mistake occurs in some papers, in particular in the recent work,\cite{55} where the authors calculated the Righi - Leduc coefficient and accepted that $  \mathcal{L} = c \cdot (\partial  M / \partial T)$ (Eq. 4 in Ref.  \cite{55}). As a result, they obtained a huge transverse heat current in crossed magnetic field and $\nabla T$,
\begin{eqnarray}
{\bf j}^h_y = - {c T\over e} \left( {d M \over dT}   \right)  \left( {d \zeta \over dT}   \right)  \nabla_x T, \nonumber
 \label{no number} 
 \end{eqnarray}
where $\zeta$ is the chemical potential. The above equation was applied to the temperature range near the superconducting transition, where the magnetization has strong temperature dependence and, therefore, the magnetization current drastically exceeds the thermomagnetic transport current. 
\par\smallskip
The examples above directly show that  all non-entropic forces should be canceled among themselves. Only in this case, the entropy flux generated by electric field corresponds to the electric current generated by the temperature gradient in accordance with Onsager relations.

\section{Gedankenexperiment in Corbino Disk Geometry}

According to Maxwell's equations, the magnetization currents are divergence-free, i.e. $ \nabla \cdot {\bf j}_{mag} = c\nabla \cdot [\nabla \times {\bf M}] = 0$. It is well understood  that as a consequence the net magnetization current that is measured in any transport experiment is always zero.\cite{42,56,57} In other words, the bulk magnetization current is always compensated by the surface currents,
\begin{eqnarray}
{\bf J}_{mag.bulk} + {\bf J}_{mag.sf} = 0. 
 \label{27} 
 \end{eqnarray}
For example, if the two-dimensional Corbino ring is placed in the perpendicular magnetic field,  the radial temperature gradient generates the circular bulk magnetization current and edge magnetization currents circulating along the inner and outer edges of the ring (see Fig. 1.A). The magnetization currents at inner and outer edges,  ${\bf J}_{mag.sf.1}$ and ${\bf J}_{mag.sf.2}$,  may be substantially different, but the net electric magnetization current,  ${\bf J}_{mag.bulk}+{\bf J}_{mag.sf.1}+{\bf J}_{mag.sf.2}$,  in any Corbino ring cross-section is always zero.\cite{42}

In the recent paper\cite{22}, the authors considered the thermomagnetic effects in the Corbino disk geometry and concluded that the surface magnetization currents provide the leading contribution to the Nernst effect. In this work, the authors assume that the inner and outer edges 
of the disk are kept at temperature $T_1$ and $T_2$, but ignore the temperature gradient inside the ring. As electrons in the bulk are not affected by the temperature gradient and the thermal entropic force given by Eq. 18 is absent, the bulk thermomagnetic current is also absent. However, this model does not corresponds to the experimental setup  for measurements of thermomagnetic coefficients. As it is highlighted in Section IV of Ref. \cite{42}, in the thermomagnetic measurements the homogeneous temperature gradient is created in a substrate and the thermal coupling to the substrate should be sufficiently strong to establish the local electron temperature in the sample to be equal to the local temperature of the substrate. In this setup the bulk and surface magnetization currents cancel each other and the thermomagnetic transport coefficient is solely determined by the electric transport current in the bulk.\cite{42} In the entropic approach, only the bulk transport current is generated by the entropic force created by temperature gradient (Eqs. 18 and 21).   
\begin{figure}[t]
\includegraphics[width=9 cm]{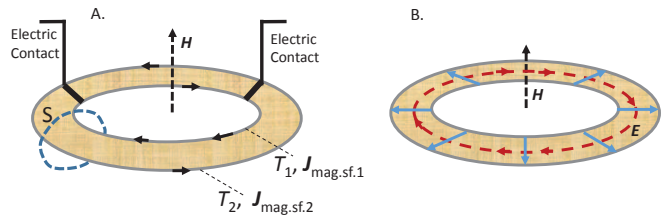}
\caption{A. Cancellation of bulk and surface electric magnetization currents induced by the radial temperature gradient in any cross-section  {\it S} of the Corbino ring.$^{42}$ B. Gedankenexperiment setup with azimuth inductive electric field and radial entropy/heat flux. There is no any surface/edge magnetization contributions to the radial current, which gives the thermomagnetic coefficient.
\label{Fig1}}
\end{figure}

While the Corbino disk geometry may be convenient for some calculations, the obtained thermomagnetic coefficient should not depend on geometry (if geometrical lengths are much larger than characteristic quantum wavelengths). Obviously, in the case of the radial temperature gradient (Fig. 1.A), the usual experimental strip geometry is obtained if the inner radius increases and becomes much larger than the width of the Corbino ring. 

Gedankenexperiment is a thermomagnetic setup with the circular inductive electric field directed along the azimuth of the Corbino geometry (see Fig. 1.B). In this case, the magnetic field perpendicular to the disc plane and the azimuth electric field generate the radial entropy/heat flux.\cite{27} As the entropy and heat are transfered in the radial direction, any surface (edge) magnetization contribution to the entropy/heat transfer is excluded in this setup.  This gedankenexperiment should give the same thermomagnetic coefficient as the usual strip setup and, therefore, we deduce that edge magnetization currents do not contribute to the entropy/heat transfer. Taking into account the divergence-free nature of magnetization currents, we conclude that bulk magnetization currents also do not transfer the entropy and heat. This consideration confirms conclusions of Sections II and III. According to magnetization thermomagnetic theory,\cite{14,15,16,17,18,19,20,21,22,29,30,31,32,33} the large heat current is transfered by the magnetization currents and, therefore, the gedankenexperiment will give the thermomagnetic coefficient strongly different from that in the strip geometry. Let us also note, that the same type of the thermomagnetic gedankenexperiment with radial electric current is impossible, because the circular temperature gradient does not exists in nature. This is a reason, why there is no analog of the Obraztsov formula (Eq. 26) for the entropy/heat transfer. 

Finally, in previous sections we discussed the Onsager relation between the bulk heat current induced by electric field and the electric transport current generated by $\nabla T$ (see Eqs. 1 and 2).  As the net electric magnetization is zero and magnetization does not contribute to the heat transfer, we see that in thermomagnetic phenomena the net electric and heat currents naturally satisfy the Onsager relation. 

\section{Contribution of Magnetization Energy to the Poynting Vector}

The study and design of metamaterials with unique electromagnetic properties has generated a lot of attention to electrodynamics of magnetic materials, in particular to the energy flow and corresponding Poynting vector. Negative phase velocity propagation and negative refraction and their interrelation have remained to be one of the most difficult aspect in the theory of metamaterials. 

 In this section we consider electrodynamics of magnetic materials and revisit the Poynting vector, ${\bf P}$, taking into account transfer of the bulk and surface electromagnetic energies related to magnetization. To simplify our consideration we assume zero dielectric polarization (the permittivity equals one)  and concentrate our attention at the magnetic contribution. According to the Maxwell equations, in a steady state the Poynting satisfy the continuity equation which expresses the energy conservation law in the following form, 
\begin{eqnarray}
\nabla \cdot {\bf P} = {\bf j} \cdot {\bf E}, 
 \label{28} 
 \end{eqnarray}
where the current ${\bf j}$ includes both transport component, ${\bf j}_{tr}$, and magnetization component, ${\bf j}_{mag}$.  It is well understood that
the term ${\bf j}_{tr} \cdot {\bf E} $ describes dissipation of electromagnetic energy into the heat. 

According to classical textbooks the magnetization does not affect the Poynting vector and in the magnetic materials it has the same form as in vacuum,\cite{56,57}
\begin{eqnarray}
{\bf P}_0 = {c\over 4 \pi} [{\bf E} \times {\bf H}].  
 \label{29} 
 \end{eqnarray}
In particular, according to Ref. \cite{56} the magnetization currents do not dissipate electromagnetic energy, do not change energy balance of electromagnetic energy, and, therefore, do not change the form of the Poynting vector. Another augment of Ref. \cite {55} in favor of Eq. 29 for magnetic materials is that the Eq.  29 ''follows independently from the obvious condition that the normal component of the Poynting vector must be continuous at the surface of a conductor, if we use the continuity of ${\bf H}_t$ and the validity of Eq. 29 in the vacuum outside the body.'' 

Two recent papers\cite{39,40} independently pointed out that in magnetic materials the tangential component of ${\bf H}$ was not continuous  at the surface due to surface (edge) magnetization currents. To take into account magnetization currents in the energy balance, the authors of Refs. \cite{39} and  \cite{40} accepted that the magnetization currents, ${\bf j}_{mag}$, contributed to the dissipation of electromagnetic energy and  produced entropy in same way as electric transport current. The dissipation increases the internal energy, 
\begin{eqnarray}
&& {\partial U \over \partial t} = {\bf j}_{tr} \cdot {\bf E} + {\bf j}_{mag} \cdot {\bf E} \nonumber \\  &&= -{c \over 4 \pi} \nabla \cdot [{\bf E} \times {\bf H}]  -c\nabla \cdot [{\bf E} \times {\bf M}]. 
 \label{30} 
 \end{eqnarray}
Then, the Poynting vector is given by\cite{39,40}
\begin{eqnarray}
{\bf P} = {c\over 4 \pi} [{\bf E} \times ({\bf H}+ 4\pi{\bf M})] ={c\over 4 \pi} [{\bf E} \times {\bf B}]  .  
 \label{31} 
 \end{eqnarray}
This form of the Poynting vector is widely accepted,\cite{58,59,60,61,62,63} while some authors arguer in favor of the classical formula (Eq. 29).\cite{41,64} However, to the best of our knowledge, the transfer of electromagnetic energy by surface magnetization currents are still overlooked in this discussion. 
\begin{figure}[t]
\includegraphics[width=9 cm]{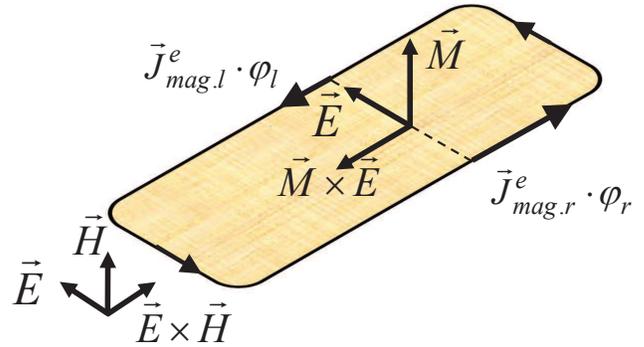}
\caption{Transfer of electromagnetic energy by bulk and surface magnetization currents
\label{Fig2}}
\end{figure}
\par\smallskip  
As we highlighted in previous sections, the quasi-stationary magnetization currents in natural magnetic materials have a quantum nature and do not dissipate energy. The same is valid for superconducting metamaterials.  Let us first neglect the imaginary part of the magnetic permeability. Without dissipation, the electromagnetic energy  may be only redistributed in the sample volume. Due to circulating character of magnetization currents,  the net electromagnetic energy transferred by magnetization through any cross-section of the sample is always zero. Thus, the electromagnetic energy transferred by the bulk magnetization currents is always compensated by the surface currents. This fundamental compensation is described  by the divergent free form of the net energy current related to magnetization,\cite{42} 
\begin{eqnarray}
{\bf j}^\epsilon_{mag} = \nabla \times
(c \phi {\bf M}) = \varphi {\bf j}^e_{mag} + c{\bf M}\times {\bf E},
 \label{32} 
\end{eqnarray}
where $\varphi$ is the electric potential. In a homogeneous sample the bulk magnetization current ${\bf j}^e_{mag} = c \nabla \times M$ is zero.  (we neglect the temperature gradient in this section). Therefore, the first term in the rhs of Eq. 32 describes the surface magnetization currents and the second term is the bulk energy flux related to magnetization. Finally, the Poynting vector, i.e. the flux of electromagnetic energy, is given by
\begin{eqnarray}
&& {\bf P} = {\bf P}_0 + {\bf P}_{mag} = {c\over 4 \pi} [{\bf E} \times {\bf H}] + c{\bf M}\times {\bf E} \nonumber \\  && = {c\over 4 \pi} [{\bf E} \times ({\bf H} - 4 \pi {\bf M})].
 \label{33} 
 \end{eqnarray}

Let us illustrate the above statements by the energy transfer in the simple two-dimensional rectangular geometry of the sample presented in Fig. 2  The surface electric magnetization current is  ${\bf J}^e_{mag.sf} = c {\bf M}\times {\bf n}$, where  ${\bf n}$ is the unit vector normal to the surface and directed
outward from the sample.  Then, the energy current transferred by the right and left edge currents is given $\varphi_{l}\cdot {\bf J}^e_{mag.l} + \varphi_{r} \cdot {\bf J}^e_{mag.r}$. Its absolute value in the direction $[{\bf M}\times{\bf E}]$ may be presented as $j^e_{mag.sf}(\varphi_{r}-\varphi_{l})=-cME\cdot w$, where $w$ is the width
of the sample. Certainly, the surface energy current is compensated by the bulk current $w\cdot c[{\bf M}\times{\bf E}]$.   

\begin{figure}[t]
\includegraphics[width=9 cm]{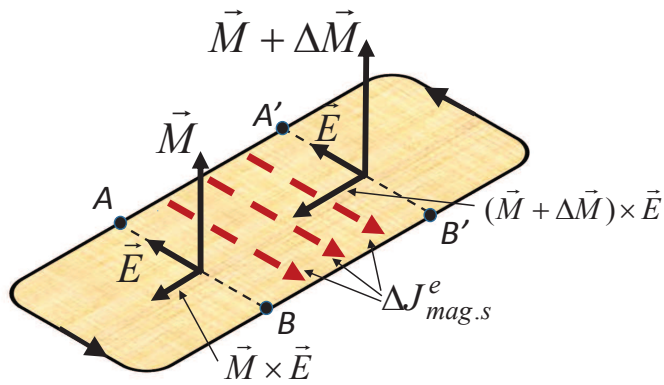}
\caption{Conservation of the electromagnetic energy in a inhomogeneous sample. The magnetization currents across the sample (red arrows) transfer the energy $ {\bf E} \cdot \Delta {\bf J}^e_{mag.s}$ from one surface to another without dissipation.
\label{Fig3}}
\end{figure}
\par\smallskip 
To clarify meaning of the term ${\bf J}_{mag}\cdot {\bf E}$ that was previously associated with dissipation,\cite{15,39,40} we consider a sample, where the permeability and corresponding magnetization ${\bf M}$ changes in the direction of ${\bf E}\times{\bf M}$ as it is shown in Fig. 3. We will analyze the energy balance in a small volume, which is formed by two close cross-sections, $(A,B)$ and $(A',B')$, shifted in the direction ${\bf E}\times{\bf M}$.  The net power carried to the volume by the bulk energy currents related to magnetization is
\begin{eqnarray}
 c w  [ ({\bf M}+ \Delta {\bf M}) \times {\bf E} ] - cw [{\bf M} \times {\bf E}]   \\ 
=  cw [ \Delta {\bf M} \times {\bf E}].  \nonumber 
\label{34} 
\end{eqnarray}
 Let us show that the incoming bulk electromagnetic energy increases the energy of the surface currents in the same volume.  First, let us note that the change of the magnetization $\Delta {\bf M}$ between $(A,B)$ and $(A',B')$ is created by the
magnetization current $\Delta J^e_{mag.s} =c \Delta M$, which flows from the left side of the sample $(A,A')$ to the right side $(B,B')$ between
the two cross-sections (A,B) and (A',B') as it is shown in Fig. 3 by the red arrows. Because of the charge conservation, the current $\Delta J^e_{mag.s} $ across the sample is exactly equal to the changes in the surface magnetization currents between $A$ and $A'$, and $B$ and $B'$: $\Delta {J}^e_{mag.s} = {J}^e_{mag.B'} - {J}^e_{mag.B} = -({J}^e_{mag.A'} -{J}^e_{mag.A})$.
The current  $\Delta {J}^e_{mag.s} = c \Delta M$ increases the surface current at right side with high electric potential and decreases the surface current at the left side with small electric potential. Thus, the energy of surface currents increases by  
\begin{eqnarray}
(\phi_B - \phi_A) \Delta J^e_{mag.s} = - {\bf E} \cdot \Delta {\bf J}^e_{mag.s} =c w |E| \Delta M, 
\label{35} 
\end{eqnarray}
where  $\Delta {\bf J}^e_{mag.s}$ is shown in Fig. 3 by red arrows. 
Compare Eqs. 34 and 35, we see that in any volume of a non-homogeneous sample, the energy conservation is reached by the redistribution between the bulk and surface energy currents. 

We limit our consideration above by the dissipationless  magnetization. Phase delay between magnetization and external magnetic field in natural magnetic materials or resistivity of metallic coils in metamaterials lead to magnetic dissipation, which is described by the imaginary part of the permeability, $\mu''$. In this case, the attenuation of the Poynting  vector is given by a factor of $\exp [-(\mu'' /\mu') {\bf k}\cdot{\bf r}]$, where ${\bf k}$ is a wavevector and $\mu'' \ll \mu'$.

Thus, the magnetization redistributes the electromagnetic energy between the bulk and surface currents. Without magnetic dissipation, the divergent free character of magnetization energy current (Eq. 32) lead to the Poyinting vector in the form of Eq.  33. For example, if $4\pi {\bf M} = {\bf H}$, i.e. the permeability $\mu =2$, the whole electromagnetic energy is transferred by the surface currents and the Poynting flux is zero. If $\mu >2 $, the energy flux transferred by surface energy current exceeds the energy current incoming from the vacuum. In this case, the Poynting flux in the bulk moves in the direction opposite to the wavevector of the incoming flux to compensate the surface energy current. 

\section{Nernst Effect of Fluctuating Cooper Pairs in Collisionless Limit}

In accordance with expectations based on the Fermi liquid picture, experimental studies did not demonstrate large thermomagnetic effects in ordinary superconductors above the superconducting transition. Therefore, experimental observations of huge Nernst  effect in various high-$T_c$ superconductors is usually considered as an evidence of non-BCS nature of superconducting transition and the non-Fermi liquid character of thermomagnetic phenomena in these materials. According to the concept proposed by Emery and Kivelson\cite{65} the low electron density and quasi-two dimensional electron spectra in superconducting cuprates lead to strong phase fluctuations and the finite temperature Berezinskii - Kosterlitz - Thouless (BKT) type transitions.  Above the transition temperature the long-range superconducting coherence is destroyed by strong phase fluctuations which lead to the generation of free vortices. The amplitude of the local superconducting order parameter remains non-vanishing in significant temperature range above $T_c$. Elementary excitations in a form of vortices determine all transport phenomena above the BKT-type transition. As it is highlighted in Ref. \cite{66} ``it is vital to recognize the elephant in the room, namely the vortex liquid above and below $T_c$.''  Also, large Nernst
signals are observed in the pseudogap state of cuprates, which is believed unrelated to vorticity.\cite{67,68} Strongly correlated pseudogap state is often associated with the holon type excitations, which originates due to mutual inelastic decay channel for electrons and holes into the charge reservoir.\cite{69} Modification of the 
of charge-conjugation symmetry can provides large thermomagnetic effects.\cite{69} Huge thermomagnetic effects based on vortices or holons do not require the Fermi liquid degeneracy factor. Surprisingly, in recent papers\cite{14,15,16,17,18,19,20,21,29} the giant Nernst effect without the $k_B T/\epsilon_F$ factor have been calculated  in the frame of traditional Gaussian fluctuation model within the Fermi liquid concept. In the next three sections we will show that giant thermomagnetic effects in the Fermi liquid without  the $k_B T/\epsilon_F$ factor are impossible.  

Here we will employ the entropic formalism of Sections II and III to calculate the fluctuation Nernst effect in the collisionless limit that provides the upper boundary for the   Nernst effect in materials with a finite electron mobility. Aronov et all.\cite{43} calculated the Hall coefficient in the collisionless limit, $\omega_c \tau \gg 1$, and  demonstrated that at $\tau \to \infty$ the fluctuation correction to the Hall coefficient vanishes, because without scattering processes it is impossible to differ the motion of electrons from the motion of Cooper pairs. In other words, in crossed electric and magnetic fields the whole electron system moves with the drift velocity, 
\begin{equation}
{\vec v}_{dr} = c \cdot {{\vec E} \times {\vec H} \over H^2}.
\label{36} 
\end{equation} 
Corrections to the entropy may be calculated from the corresponding corrections to electron free energy in the fluctuation region, $S= - \partial F/ \partial T$. For two dimensional superconductor ($d \ll \xi $, $d$ is the conductor thickness, and $\xi $ is the coherence length in pure superconductor) the singular part of the fluctuation free energy is\cite{15}
\begin{equation}
 F_{fl} (\epsilon, h) = - {T \ h \over 2 \pi \xi^2 } \cdot 
 \ln {\Gamma \left( {1 \over2} + {\epsilon \over 2h} \right) \over \sqrt {2 \pi} }, 
\label{37} 
\end{equation}
where dimensionless parameters $\epsilon = 1 - T/T_c$ and $ h= H/ H_{c2} (T)$ are described temperature and magnetic field dependency of the free energy. Thus, the singular part of the entropy is
\begin{equation}
 S_{fl} = - \begin{cases} {1 \over 4 \pi \xi^2} \cdot \ln\epsilon ,& h \ll \epsilon \ll 1;  \\
{\ln2 \over 4 \pi^2 r_H^2} ,&  \epsilon\ll h \ll 1;  \\
{1\over 2 \pi \xi^2} \cdot {h \over \epsilon +h} ,&    \epsilon +h \ll h \ll 1.   \end{cases} 
\label{38} 
\end{equation}
Finally, we can calculate the heat current, ${\vec J}^h = TS_{fl} {\vec v}_{dr}$,  and determine the thermomagnetic coefficient, 
\begin{equation}
\mathcal{L}= - {1\over 2} \beta_0 \begin{cases} \left({r_H \over \xi}\right)^2 \ln {T \over T-T_c} ,& h \ll \epsilon \ll 1;  \\
 \ln2 ,&  \epsilon\ll h \ll 1;  \\
  \left({r_H \over \xi}\right)^2 {H_{c2} (T) \over H-H_{c2}(T)} ,&    \epsilon +h \ll h \ll 1,   \end{cases} 
\label{39} 
\end{equation}
where $\beta_0 = ke/\pi \hbar $=6.68 nA/K is the quantum of thermoelectric conductance. 

Taking into account that $\xi \sim \hbar v_F/k_B T_c $, we see that in small magnetic field the fluctuation correction to the electronic entropy (Eq. 38) is of the order of $kT_B/\epsilon_F$ with respect to the normal state entropy, which is $\sim ( p_F/\hbar)^2 k_B T/\epsilon_F$. Therefore, the entropy related to superconducting fluctuations cannot provide the giant Nernst effect.   

\section{Nernst Effect of Fluctuating Cooper Pairs: Microscopic Calculations in Weak Magnetic Field}

Theoretical and experimental investigations of the Nernst effect due to fluctuating Cooper pairs above the superconducting transition temperature, $T_c$, has a long history. As we discussed in previous sections, in normal metals in a weak magnetic field the Nernst effect is proportional to $(\omega_c \tau)\cdot (k_B T/\epsilon_F)$ and, therefore, it is very small. Considering the Nernst effect, as the Seebeck effect in magnetic field or as a thermal Hall effect, one can say that the Nernst coefficient in normal metals accumulates both the entropy related degeneracy factor of the Seebeck coefficient and the particle - hole asymmetry (PHA) factor of the Hall effect related to time-reversal asymmetry in magnetic field.  As was shown by Fukuyama et all.\cite{70}, for the  fluctuating Cooper pairs both the Hall and Seebeck coefficients are proportional to small PHA term in the fluctuation propagator. These results were confirmed by dozens of experimental investigations in traditional and high-T$_c$ superconductors.\cite{15}  The Nernst coefficient due to fluctuations was calculated by Maki\cite{29} employing the Kubo method.  Surprisingly, Maki obtained the giant Nernst effect (GNE),\cite{29} which does not require any PHA and, therefore, it $(\epsilon_F/k_B T)^2 \sim 10^8-10^{10}$ times larger then the Nernst effect in normal metals. Ten years later, Ullah and Dorsey obtained the same results using phenomenological time dependent Ginzburg Landau equation.\cite{30} After discovery of  high-T$_c$ superconductors, several theoretical groups\cite{14,15,18,19,31,32,33,34,35,36}  reproduced the giant Nernst effect due to fluctuating Cooper pairs to explain the experimental data  in high-T$_c$ materials without appealing to vorticity. 

GNE due to fluctuating Cooper pairs also looks rather controversial, if one tries to find excitations which can transfer giant entropy that is required by GNE. GNE has been challenged in our work.\cite{27} Employing the gauge-invariant form of the electron heat current in magnetic field, we show  that the Nernst effect due to fluctuating Cooper pairs in weak magnetic field is proportional to the square of PHA. Controversy related to the fluctuation Nernst effect was discussed in the comments,\cite{28,71} but no agreement has been reached. 
 
There are two main problems in the theory of the Nernst effect of fluctuating Cooper pairs. The first one is the general problem of the gauge invariance of the Kubo method for superconductors in magnetic field. The second problem is the microscopic form of the heat current operator of interacting electrons. In general,  the heat current operator is obtained from the Luttinger effective gravitational field\cite{72} or from the energy-momentum tensor, which is calculated from the Lagrangian of the system (for details see Ref. \cite{52}).  Keeping time derivative of the electron field operators in the energy-momentum tensor, one obtains the heat current operator in the ``$\epsilon$-representation'', i.e. in the frequency domain. Employing the equation of motion, the time derivative of field operators may be expressed via the electron field operators to get the ``$\xi$ - representation'' of the heat current, i.e. the heat current in the energy domain.  The $\xi$- representation is physically transparent, as its every term may be directly associated with the kinetic energy or the interaction energy, but it leads to a large number of diagrams to be considered.\cite{52} For the fluctuation thermomagnetic effects, both representation of the electron heat current provide consistent results.\cite{15,51} For the fluctuation Nernst effect, In Refs. \cite{73,74} and \cite{75} we employed the the gauge invariant $\xi$-representation, while in other works\cite{14,15,18,19,32,33} the heat current was used in the $\epsilon$-representation.  

To finally resolve the  long-standing controversy related to the Nernst effect of fluctuating pairs, in this section we will employ the well-established gauge-invariant linear response formalism developed by Aronov, Hikami, and Larkin for the fluctuation Hall coefficient.\cite{43} Implementing the heat current operator in the same form as it was used in Refs. \cite{14,15,18,19,32}  and \cite{33} that obtained GNE, we will show that in the gauge invariant formalism the giant Nernst effect of fluctuating Cooper pairs, i.e. the Nernst effect without PHA,  does not exist. 

It is well established that in the Kubo method the magnetization currents do not contribute to the Hall conductivity,  $\sigma_{xy}$, which is given by 
\begin{eqnarray}
 \sigma_{xy} \equiv \lim_{\Omega \rightarrow 0} \  \langle [(j^e_{tr})_x  (j^e_{tr})_y] \rangle = \lim_{\Omega \rightarrow 0} \  \langle [(j^e_{bulk})_x  (j^e_{bulk})_y] \rangle, 
 \label{40} 
 \end{eqnarray}
where  $ \langle [j^e_x  j^e_y] $ is the Fourier representation of the retarded correlation functions of electric currents $J^e_x$ and and $J^e_y$. 

\begin{figure}[t]
\includegraphics[width=9 cm]{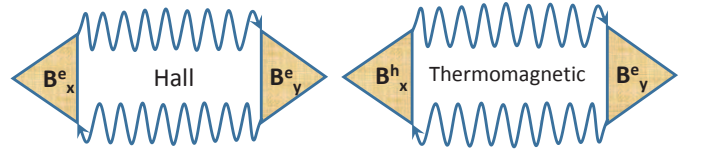}
\caption{Kubo diagrams for the Hall coefficient and thermomagnetic coefficient of fluctuating Cooper pairs. The wavy line is for the fluctuation propagator  $L_{n}$ (Eq. 53), the $B^e$-triangle is the electric current operator, and the $B^h$-triangle is the heat current operator for Cooper pairs.
\label{Fig4}}
\end{figure}

\par\smallskip 
The thermomagnetic coefficient, $\mathcal{L}_{xy}$, may be presented in analogous way, 
 \begin{eqnarray}
\mathcal{L}_{xy} \equiv \lim_{\Omega \rightarrow 0} \ {1\over T}\langle [j^h_x  (j^e_{tr})_y] \rangle = \lim_{\Omega \rightarrow 0} \ {1\over T}\langle [j^h_x  (j^e_{bulk})_y] \rangle, 
 \label{41} 
 \end{eqnarray}
 where $\langle [j^h_x \ j^e_y] \rangle$ is the Fourier representation of the retarded correlation functions of the heat current $J^h_x$ and the electric current $J^e_y$. Magnetization currents do not contribute to the heat current, $j^h$, and in the linear response the magnetization currents do not correlate with the heat transfer. Therefore, in the correlation function $\langle [j^h_x  (j^e_{tr})_y]$  the electric transport current $j^e_{tr}$ may be replaced by the bulk electric current $j^e_{bulk}$.  
Thus, the thermomagnetic coefficient given by the Kubo formula does not require any correction from magnetization currents.  

To calculate the thermomagnetic coefficient, $ \mathcal{L}_{xy}$, in the Kubo method one should replace the electric current operator, $B^e_x$, in the equation for the Hall coefficient, $\sigma_{xy}$, by the heat current operator $B^h_x$ (see Fig. 4):
\begin{eqnarray}
 \sigma_{xy} = \lim_{\Omega \rightarrow 0} \ \int {d\omega \over \Omega} \ B^e_x B^e_y  F(\omega, \Omega), \\
 \mathcal{L}_{xy} = \lim_{\Omega \rightarrow 0} \ {1\over T} \int {d\omega \over \Omega} \ B^h_x B^e_y  F(\omega, \Omega), 
\label{42,43} 
\end{eqnarray}

The heat current vertex for fluctuating Cooper pairs is usually taken in the form,\cite{14,15,18,19,32,33}
\begin{equation}
B^h (\omega) = {\omega \over 2e} B^e. 
\label{44} 
\end{equation}
Let us also note, that some authors\cite{33} together with $\omega$ also include the Kubo frequency, $\Omega$, to the heat current vertex. First of all, appearance of the Kubo frequency in the thermal energy is erroneous, because the thermal energy is defined via the entropy, which is the function of the probability distribution in the statistical ensemble and it does not fluctuate since it is not a function of the state in which the system happens to be.  Nevertheless, even if  $\Omega$ is erroneously included in the thermal energy, this term does not provide any contribution to the Nernst coefficient (Eq. 43), because $F(\omega, \Omega=0) = 0$ in accordance with the causality principle for the Hall effect (Eq. 42). Thus, to calculate the thermomagnetic coefficient (Eq. 43) one should {\it just add  $\omega /2eT $ to the expression for the Hall coefficient}  (Eq. 42), 
\begin{equation}
 \mathcal{L}_{xy} = \lim_{\Omega \rightarrow 0} \ {1\over 2e T} \int {d\omega \over \Omega} \ \omega  B^e_x B^e_y  F(\omega, \Omega). 
 \label{45} 
\end{equation} 
  
The most accurate way to obtain the gauge invariant expression for the Hall coefficient is to apply the Kubo method to electron states in Landau representation, which directly includes the magnetic field. For fluctuating Cooper pairs this approach has been developed in Ref. \cite{43}. According to Eqs. 6 and 13 of Ref. \cite{43}, the Hall coefficient in a two dimensional superconductor in Landau representation is given by   
\begin{eqnarray}
 && \sigma_{xy} = - \lim_{\Omega \rightarrow 0} {(4e \xi_\ell^2)^2\over 8 \pi^2r_H^4} \sum_{n=0}^{\infty} (n+1) 
 \nonumber \\  &&
 \times \int \limits_{-\infty}^{\infty} {d \omega \over \Omega}  \coth{\omega \over 2T}   \Phi (n, \omega, \Omega),    
\label{46} 
\end{eqnarray}
 where $\Phi (n, \omega, \Omega)$ is given by  
 \begin{eqnarray} 
&&  \Phi (n, \omega, \Omega)  = \Im L_{n}^R (\omega) \cdot \left[ L_{n+1}^R(\omega+\Omega) - L^A_{n+1}(\omega-\Omega) \right] 
\nonumber \\ 
&&  - \ \Im L_{n+1}^R(\omega)  \cdot \left[ L_{n}^R (\omega+\Omega) - L_{n}^A (\omega-\Omega) \right] , 
\label{47} 
\end{eqnarray}
where  
$\xi_\ell$ is the coherence length in 2D disordered superconductor, $r_H^2=(2eH)^{-1}$, and $L_{n}^R$ is the fluctuation propagator in the Landau representation. In the zero order in PHA, the fluctuation propagator is  
 \begin{equation}
  L_{n}^{R(A)} = \left( {T-T_c \over T_c} \mp  {i \pi \omega \over 8 T_c} +{\xi_\ell^2 \over r_H^2}(n+1) \right)^{-1}.
\label{48} 
\end{equation}

Calculating the linear response to applied electric field, one should keep in Eq. 46 only the linear terms in $\Omega$.
Also, in a weak magnetic field ($r_H \rightarrow \infty $), we may expand the integrand $ \Phi(n, \omega, \Omega)$ in power of $1/n$ and take into  account only the first term in this expansion. In the first order in $\Omega$ and $1/n$, we get 
\begin{eqnarray}
 \Phi (\omega, \Omega) =\Omega \cdot {8 \over 3} {\xi_\ell^2 \over r_H^2} \cdot { \partial \left( \Im L^R (\omega)\right)^3 \over \partial \omega},  
\label{49} 
\end{eqnarray}
where the fluctuation propagator in a weak field is 
\begin{equation}
  L^{R(A)}(\omega, q) = \left( {T-T_c \over T_c} \mp {i \pi \omega \over 8 T_c} +\xi_\ell^2 q^2  \right)^{-1}. 
\label{50} 
\end{equation}
Substituting the summation over $n$ by the integration over $q$ in Eq. 46 we obtained the final equation the Hall conductivity of fluctuating Cooper pairs in a weak magnetic field,   
\begin{eqnarray}
&& \sigma_{xy} = - {2 (4e \xi_\ell^2)^2 \xi_\ell^2 \over 3 \pi r_H^2} \int_0^{\infty} q^2 {2 \pi q dq \over (2\pi)^2} 
\nonumber \\  && \times \int_{-\infty}^{\infty} d \omega \ \coth{\omega \over 2T} \cdot { \partial \left( \Im L^R (\omega, q)\right)^3 \over \partial \omega},  
\label{51} 
\end{eqnarray}
Let us note that integrating Eq. 51 over $\omega$ by parts we will get the Eq. 14 of Ref. \cite{43}.  As $\Im L^R(- \omega) = -\Im L^R(\omega)$,  the Hall conductivity is zero in zero order in PHA. To get the Hall conductivity in the first order in PHA, one should add the PHA term,  $ \gamma \omega/ 2\epsilon_F $, to the fluctuation propagator,\cite{43,70}  
\begin{equation}
  L_{n}^{R(A)} = \left( {T-T_c \over T_c} \mp  {i \pi \omega \over 8 T_c} + {\gamma\omega \over 2 \epsilon_F}+{\xi_\ell^2 \over r_H^2}(n+1) \right)^{-1}. 
 \label{52} 
\end{equation}
where $\gamma = \partial \ln T_c /\partial \ln \mu$, and $\mu$ is the electron chemical potential.\cite{43}
Finally,  the fluctuation Hall coefficient is given by\cite{43}
 \begin{eqnarray}
\sigma^{fl}_{H} = {e \gamma \over 48} \cdot \omega_c \tau   \cdot \left( {T_c \over T-T_c} \right)^2.
\label{53}
\end{eqnarray}

Replacing the electric current operator by the heat current operator in Eq. 46, we obtained the thermomagnetic coefficient, 
\begin{eqnarray}
&& \mathcal{L}^{fl}  = - \lim_{\Omega \rightarrow 0} {(4e \xi_\ell^2)^2\over 8 \pi^2r_H^4} \sum_{n=0}^{\infty} (n+1)   \nonumber \\  && \times \int \limits_{-\infty}^{\infty} {d \omega \over \Omega}  \ \omega \coth{\omega \over 2T}   \Phi (n, \omega, \Omega),    
\label{54} 
\end{eqnarray}
 where $\Phi (n, \omega, \Omega)$ is given by  Eq. 47, which is the same as for the Hall coefficient. Therefore, in a weak magnetic field in zero order in PHA (Eqs. 48 -50) we obtained 
\begin{eqnarray}
&& \mathcal{L}^{fl} = - { e (4 \xi_\ell^2)^2 \xi_\ell^2 \over 3 \pi r_H^2} \int_0^{\infty} {q^3 dq \over (2\pi)} 
 \nonumber \\  && \times \int_{-\infty}^{\infty} d \omega \ \omega \coth{\omega \over 2T} \cdot { \partial \left( \Im L^R (\omega, q)\right)^3 \over \partial \omega},  
\label{55} 
\end{eqnarray}
The heat current vertex brings additional $\omega$, which changes the parity of the integrant for the thermomagnetic coefficient (Eq. 55) with respect to the integrant for the Hall coefficient (Eq. 51). According to Refs. {18,19}, this leads to giant thermomagnetic coefficient in zero order in PHA.  However, one needs to be careful with  this conclusion. Note that the function $\Phi(\omega)$ is the full derivative (see Eq. 49) and, therefore, integrating Eq. 55 over $\omega$ by parts, we get
\begin{eqnarray}
&& \mathcal{L}^{fl} = { e (4 \xi_\ell^2)^2 \xi_\ell^2 \over 3 \pi r_H^2} \int_0^{\infty} {q^3 dq \over (2\pi)} 
 \nonumber \\  && \times \int_{\infty}^{\infty} d \omega \    \left( \Im L^R (\omega, q)\right)^3  \cdot { \partial  \over \partial \omega}  \left( \omega \coth{\omega \over 2T}\right).  
\label{56} 
\end{eqnarray}
The only essential difference of Eq. 56 from the corresponding equation for the Hall coefficient given by Eq. 51 (Eg. 14 in AHL paper)  is the presence of an additional $\omega$, which appears in Eq. 56 in combination with the distribution function, $\coth (\omega/2T)$. As the characteristic range of $\omega$ is $\sim T-T_c \ll T_c$, the derivative of $\omega \coth(\omega / 2T)$ over $\omega$  is $\omega /3T \ll 1$. For this reason, fluctuations of Cooper pairs do not provide any singular contribution to the thermomagnetic coefficient. 

To obtain singular contribution to the Nernst coefficient, one should take into account PHA term in the fluctuation propagator (Eq. 52). Calculating the Nernst coefficient (Eq. 54) we should expand the fluctuation propagator up to the second order in PHA (the first order contribution is zero due to the parity). In this way we reproduce our result in Ref. \cite{27} obtained with the $\xi$ - presentation of the heat current operator, 
\begin{eqnarray}
&& \mathcal{L}^{fl} = - {5\gamma^2 \over 4} \beta_0 \left({\xi_\ell \over r_H}\right)^2   \left({4  T_c \over \pi \epsilon_F} \right)^2  { T_c \over T-T_c} 
\nonumber \\  && = - {5 \gamma^2  \over \pi} \beta_0 \cdot \omega_c \tau \cdot  {T_c \over \epsilon_F } \cdot { T_c \over T-T_c} 
\label{57} 
\end{eqnarray}
where $\beta_0= k_B e/\pi \hbar = 6.68$nA/K is the quantum
of thermoelectric conductance.  In Ref. \cite {27} the fluctuation thermomagnetic coefficient was calculated by two different methods. In the Kubo method, we employ the gauge - invariant  form of the heat current operator in the $\xi$-representation, which was derived from the energy-momentum tensor (Refs. \cite{73} and \cite{74}) and the quantum energy balance equation (Ref. \cite{75}).  Using the quantum transport equation in the Keldysh formalism, we calculated both the heat current generated by the electric field and the electric current induced by temperature gradient. 

Summarizing our calculations of the Nernst coefficient of fluctuating Cooper pairs, we would like to highlight mathematical and physical rationales for the absence of giant Nernst effect, which was predicted in a number of theoretical papers in frames of BSC model,\cite{14,15,18,19,31,32,33} but has never been observed in traditional superconductors. Mathematically, this is a result of the form of the integrands  in Eqs. 51 and 54. The only dereference between the integrands is the appearance of $\omega$ in the thermomagnetic coefficient. Let us note that the thermal energy of Cooper pairs, $\omega$, always appears in combination with the distribution function $\coth(\omega / 2T)$ (Eq. 56) and never appears in combination with its derivative. Close to $T_c$, the function $(\omega/2T) \coth (\omega/2T)$ is $\sim 1$ and, therefore, it does not provide any singular contribution to the thermomagnetic coefficient without taking into account PHA in the fluctuation propagators. 

According to Eq. 57, the fluctuation thermomagnetic coefficient scales as the product of  $\omega_c \tau$  and $T/\epsilon_F$, i.e. exactly in the same way as the the normal state thermomagnetic coefficient. Let us note, that the fluctuation Hall coefficient (Eq. 53) scales as $\omega_c\tau$, i.e. as the Hall coefficient in the normal state. Fluctuation thermoelectric coefficient  scales as $T_c / \epsilon_F$, i.e. as the thermoelectric coefficient in the normal state.\cite{15,51} Therefore, the scaling  $\omega_c \tau \cdot T/\epsilon_F$ is highly expected for the fluctuation thermomagnetic coefficient. The characteristic thermal energy transfered by fluctuation Cooper pairs, $\omega$, is of the order of $T - T_c$. Therefore, the fluctuation thermoelectric power is given by the fluctuation conductivity multiplied by the product of $T_c/\epsilon_F$ and $(T-T_c)$. In the same way, 
the thermomagnetic coefficient (Eq. 57) is given by the fluctuation Hall coefficient (Eq. 53) multiplied by the product of $T_c/\epsilon_F$ and $(T-T_c)$. The giant thermomagnetic coefficient obtained in  Refs. \cite{14,15,18,19,31,32} and \cite{33} is by a factor of $(\epsilon_F/T_c)^2$ larger than that given by Eq. 57. 

Thus, the fluctuating Cooper pairs cannot give the giant Nernst effect because the Cooper pairs do not transfer the entropy. The only possible effect of superconducting pairing is the reduction of the entropy of normal electrons. Therefore, the expected thermomagnetic coefficient in the fluctuation region cannot exceed the thermomagnetic effect of normal electrons, which is proportional to the square of PHA and small.  Giant thermomagnetic effects may be associated with vortices and will be discussed in Section IX.  

\section{Thermomagnetic Entropy per Charge}

In Sections II and III we consider thermomagnetic effects in the collisionless limit in terms of entropy fluxes and entropy - related thermal forces.
This approach provides clear physical picture, which is not obscured by various temperature - dependent properties of materials and structures that are not related to the entropy.  In this Section we generalize the entropic approach to the disordered conductors with interacting (correlated) electrons.

Again, we divide the electron system into a large number of small statistical subsytems of charge carriers and calculate the linear response of these subsytems to the crossed temperature gradient and magnetic field. The thermal force per electron in the $i$-subsytem, ${\bf f}^{th}_{ i}$ (Eq. 18), leads to the drift velocity of a carrier, ${\bf v}^\parallel_ i = \mu^\parallel_i {\bf f}^{th}_{ i} $, where $\mu^\parallel_i $ is the electron mobility in the direction of the temperature gradient.
In the direction perpendicular to the temperature gradient,  the Lorentz force, ${\bf f}^{L}_{ i}$ provides the electron drift, ${\bf v}^\perp_ i = \mu^\perp_i {\bf f}^{L}_{ i} $. Calculating corresponding thermomagnetic current, we obtained  the thermomagnetic coefficient,
\begin{eqnarray}
\mathcal{L}  = - {H \over c} \sum_i \mu^\parallel_i  \mu^\perp_i \Delta S_i 
= - {H \over c} \sum_i \mu^\parallel_i  \mu^\perp_i \tilde{s}_i \Delta N_i,
\label{58}
\end{eqnarray}
where ${\tilde s}_i = \Delta S_i / \Delta N_i$ is the entropy per carrier in the $i$-group of charge carriers. Eq. 58 reproduces the thermomagnetic coefficient for noninteracting electrons. 

In particular, in an isotropic conductor with the energy - independent  electron momentum relaxation time, $\tau$, we obtain $\mathcal{L} =- H  \mu^2 S_e /c $, where $\mu = | e | \tau /m$.  Taking into account the thermodynamic relation between the entropy per electron and the temperature derivative of the electron chemical potential, $S_e/n_e = (\partial \zeta / \partial T)_{n_e}$, the thermomagnetic coefficient in this case is given by 
 \begin{eqnarray}
\mathcal{L} =- {H \over c} \mu^2 S_e = - {H \over c} \mu^2 n_e  \left( {\partial \zeta \over \partial T} \right)_{n_e}.
\label{59}
\end{eqnarray}

Let us note, that the same-type formula, but with full temperature derivative was proposed by Varlamov et all.\cite{34,35,36} directly for the Nernst coefficient $\mathcal N$, given by Eq. 3. Our consistent derivation shows that applicability of Eq. 59 is rather limited  by the isotropic and energy-independent electron mobilities. These limitations do not allow for account the delicate balance of all transport coefficients in the equation for the Nernst effect (Eq. 3). In particular, according to Sondheimer\cite{4}, if the momentum relaxation time does not depend on electron energy, the Nernst coefficient is zero. 

Let us also note, that Eq. 59 presents thermomagnetic coefficient via the thermodynamic entropy of electrons or, alternately, via the temperature derivative of the thermodynamic chemical potential.  In some papers\cite{34,35,36}  the giant fluctuation Nernst coefficient is explained by the strong temperature-dependence of the chemical potential of Cooper pairs.  Like the chemical potential of the photons in equilibrium black-body radiation, the thermodynamic chemical potential of Cooper pairs is always zero due to the lack of constraint on the number of Cooper pairs in the system.\cite{23}  

As the Nernst effect is described in terms of thermoelectric and thermomagnetic transport coefficients, let us present these coefficients in terms of the entropy.  It is well understood that the thermoelectric Seebeck coefficient is given by the entropy per unit charge, 
\begin{eqnarray}
\mathcal{S}  = { \bar{s}_{th-e} \over |e|} , 
\label{60}
\end{eqnarray}
where the entropy  per carrier charge, $ \bar{s}_{th-e}/|e|$,  is defined as a ratio of the entropy current to the electric current induced by the same electric field. 
In terms of electron mobility, the the entropy  per charge is given by 
\begin{eqnarray}
{\bar{s}_{th-e} \over |e|} =   { \sum_i   q_i  \tilde{s}_i   \mu^\parallel_i  \Delta N_i \over  e^2 \sum_i    \mu^\parallel_i  \Delta N_i }, 
\label{61}
\end{eqnarray}
where $q_i = \mp |e|$ for electrons and holes, correspondingly.  

In analogy with the thermoelectric phenomena, it will be convenient to introduce the average thermomagnetic entropy per unit charge, $ \bar{s}_{th-m}$,  that is defined as a ratio of transverse entropy current to the transverse electric current induced  by crossed electric and magnetic fields. 
Then transverse transport coefficients are interrelated as 
\begin{eqnarray}
{\mathcal{L} \over \sigma_H} = - {\bar{s}_{th-m}  \over |e|}. 
\label{62}
\end{eqnarray}
In accordance with Eq. 58, in the disordered limit, $\omega_c \tau \ll 1$, the thermomagnetic entropy per unit charge is 
\begin{eqnarray}
{\bar{s}_{th-m} \over |e|} =   { \sum_i    \tilde{s}_i  \cdot \mu^\parallel_i \mu^\perp_i  \Delta N_i \over   \sum_i  q_i  \mu^\parallel_i \mu^\perp_i  \Delta N_i }. 
\label{63}
\end{eqnarray}
 In the opposite, collisionless limit, $\omega_c \tau \gg 1$, the whole electron system drifts with the same velocity and the entropy per charge is 
\begin{eqnarray}
{\bar{s}_{th-m} \over |e| } =   {  S_e \over |e| N_e }, 
\label{64}
\end{eqnarray}
where $S_e$  is the total electron entropy and $N_e$ is the number of electrons. 

Finally, in accordance with Eq. 3, the measured Nernst coefficient is given by 
\begin{eqnarray}
 \mathcal{N} =  \left({\bar{s}_{th-e} \over |e|} - { \bar{s}_{th-m} \over |e|} \right)  \cdot {1 \over H}\cdot { \sigma \cdot \sigma_H \over \sigma^2 +\sigma_H^2}. 
\label{65}  
\end{eqnarray}
Thus, the Nernst coefficient is proportional to the difference of thermoelectric and thermomagnetic entropies per unit charge. If the electron momentum relaxation time is independent on energy and momentum direction, both thermoelectric and thermomagnetic entropies per unit charge equal $S_e /|e| N_e$ (Eqs. 61 and 63), and the Nernst coefficient is zero. This explains the Sondheimer cancellation in terms of the entropy. Also, in weak disorder (strong magnetic field) the thermoelectric entropy per charge is also given by $S_e /|e| N_e$ (see Ref. \cite{42}) and, according to Eqs. 64 and 65, the Nernst coefficient is zero. Large Nernst effect requires large entropy of some group of electrons, $\tilde{s}_i $, as well as substantially different averaging procedures for thermoelectric (Eq. 61) and thermomagnetic (Eqs. 63) entropies. Significant difference between thermoelectric and thermomagnetic entropies per carrier charge is also expected for conductors with strongly anisotropic mobility.  The consideration above shows that in the Fermi liquid the thermomagnetic coefficient is proportional to the entropy per carrier charge, which is always proportional to $k_B T/\epsilon_F$. The Nernst coefficient may be further reduced due to small difference between thermoelectric and thermomagnetic entropies per unit charge. 

\section{ ${\bf \nabla T}$ - induced Transverse Vortex Transport}

\subsection{Nonentropic Forces Generated by Moving Temperature Gradient}
Thermomagnetic effects have been intensively studied in vortex state of traditional and high-$T_c$ superconductors.\cite{7,49,50,76,77} 
Superconducting vortex consists of a normal core with the size of the coherent length, $\xi$, and superconducting currents circulating around the core in the area of the order of the magnetic penetration length, $\lambda$. Both the normal core and superconducting currents contribute to the free energy of the vortex. These contributions with respect to the superconducting background per a unit of vortex line are given by\cite{7}
\begin{eqnarray}
&&F_\phi^n (T) = {H_c^2 \over 8 \pi} \pi \xi^2 = \left( {\Phi_0 \over 4 \pi \lambda } \right)^2, \\
&&F_\phi^s (T) = {H_c^2 \over 8 \pi} \pi \xi^2 \ln (\lambda/ \xi) = \left( {\Phi_0 \over 4 \pi \lambda } \right)^2  \ln (\lambda/ \xi) ,
\label{66, 67}
\end{eqnarray}
where $H_c^2 (T)$ is the critical magnetic field, and $\Phi_0$ is  the quantum of the magnetic flux. The free energy of superconducting currents is logarithmically exceeds the free energy of the core and, therefore, the core contribution is usially neglected.\cite{7,49,50,76} 

In the Nernst effect, vortices under the thermal force move in the direction of $\nabla T$ and transfer the magnetic flux, which generates the transverse Nernst voltage. In the Ettingshausen effect, vortices under the Lorentz force move perpendicular to the electric field ( electric current) and transfer the thermal energy in this direction.\cite{7,49,50,76,77} In the original work by Stephen\cite{49} and other papers the leading term in the transport entropy was associated with the superconducting currents (Eq. 67). The vortex entropy was revisited in our recent paper,\cite{26} where we have highlighted that the thermodynamic entropy should be calculated at constant magnetization (Eq. 9) and, therefore, the entropy of superconducting currents is zero. This is also highly expected from the quantum nature of superconducting currents. Besides the configuration entropy of vortices, which is usually small (see next subsection), the only entropy that appears in vortex thermomagnetic phenomena is the entropy of a vortex core. Both the thermal force and the heat current are described in terms of the transport entropy of a vortex, $S_d$,\cite{26} 
\begin{eqnarray}
S_d&=&  - {\partial F_\phi^n (T) \over \partial  T} , \\
 {\bf f}_{th} &=& - S_d \nabla T 
=  - \left( {\Phi_0 \over 4 \pi } \right)^2  {1\over (\lambda(T))^3}  
{\partial \lambda(T) \over \partial T}   \nabla T, \\
 {\bf j}^h &=& TS_d {\bf v}_{dr}. 
\label{68,69,70}
\end{eqnarray}

\begin{figure}[t]
\includegraphics[width=9 cm]{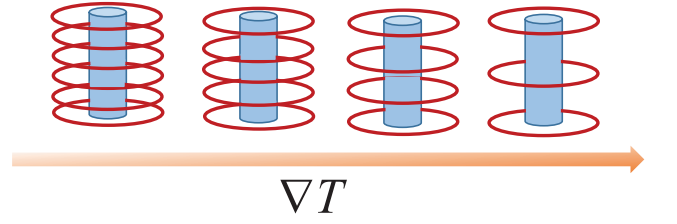}
\caption{Single vortex moves from the cold to the hot area to minimize the kinetic and electromagnetic energy of superconducting magnetization currents circulating around the vortex core. This effect does not contradict to thermodynamics, because the magnetization currents  do not transfer entropy. 
\label{Fig.5}}
\end{figure}

Supercurrents around cores neither produce the thermal force, nor participate in the heat transport. For this reason, the 
thermomagnetic effects are negligible in a system of Josephson vortices with a insulator core in a superconductor-insulator-superconductor junctions.\cite{78} At the same time, the temperature dependence of the free energy of supercurrents, $F_\phi^s (T)$, leads to the non-entropic force, 
\begin{eqnarray}
{\bf f}_{s} = - {\partial F_\phi^s (T) \over \partial {\bf r}}  
 =  \left( {\Phi_0 \over 4 \pi } \right)^2  {1\over (\lambda(T))^3}  
{\partial \lambda(T) \over \partial T} \ln {\lambda\over  \xi} \cdot \nabla T. 
\label{71}
\end{eqnarray}
As the energy of supercurrents decreases with temperature, this non-entropic force drives a vortex from the cold to the hot area. Moreover,
the non-entropic force (Eq. 67) exceed the thermal force related to the core entropy by a large logarithmic factor. 
Of course, in quasi-equilibrium conditions, i.e. in the homogeneous temperature gradient, such motion of vortices contradicts the second  thermodynamic law. In Ref. \cite{26}, we have shown that the non-entropic force given by Eq. 71 is balanced by another non-entropic force, which is the Lorentz force between the vortex and superconducting magnetization currents created by $\nabla T$.\cite{26}  Let us note that the Lorentz force may be also described as vortex repulsion from the hot area with high vortex concentration to the cold area with low vortex concentration. Due to fundamental relation between the electromagnetic energy and magnetization of interacting vortices,\cite{78} this cancellation of ${\bf f}_{s}$ and the Lorentz force is also valid for strongly interacting and even overlapping vortices in the homogeneous temperature gradient.  However, the local or transient temperature gradient does not provide vortex redistribution, does not create corresponding magnetization currents, and does not generate the Lorentz force. In this case, the non-entropic force related to the temperature dependent superconducting currents drives a single vortex in the direction of $\nabla T$.

Such vortex motion has been demonstrated in recent unique experiments with single superconducting vortices that follow a laser-generated hot spot.\cite{44} The vortex motion was investigated in superconducting niobium films with a thickness of 90 nm
grown by magnetron sputtering on a silicon substrate. For imaging of individual vortices the authors develop a magneto-optical system based on 
the Faraday rotation of light polarization in a magneto-optical indicator placed onto the superconductor, in a crossed-polarizer beam path configuration. Ref \cite{44} has demonstrated precise manipulation of individual vortices by the laser-induced non-entropic force. 

Obviously, the static laser-generated hot spot creates classical thermomagnetic phenomena, where vortices move from the hot to cold area. If  the laser hot spot moves with the velocity below the speed of vortices, the vortices will be accumulated in the hot sport. Sooner or later the vortex concentration in the hot sport will reach the quasi-equilibrium value that corresponds to the hot spot temperature and the many-body repulsive force will equilibrate the force ${\bf f}_{s}$. 

Let us evaluate the minimal velocity of the hot sport to observe the vortex movement from the cold area to the hot area. This velocity equals to the vortex driving speed, which is given by the Stephen-Bardeen vortex viscosity, $\eta_{S-B}$,  
\begin{eqnarray}
{\bf v}_{\phi} = {{\bf f}_{s} \over \eta_{S-B}}, \  \  \  \  \eta_{S-B}= {\pi \hbar^2 \over 4 e^2 \xi^2 \rho}
\label{72}
\end{eqnarray} 
where  $\rho$ is the film resistivity. In particular, for experimental conditions in Ref. \cite{44}, - $T$ = 4.6 K and $\nabla T \sim$ 1K/$\mu$m, -
we evaluate the vortex driving speed as 10 $\mu$m/s, which is in a good agreement with the speed experimentally observed in Ref. \cite{44}.
Driving speeds of vortices in electric fields are significantly higher.\cite{80} The minimal scanning speed, i.e. velocity of the hot  spot, depends on the laser power. High power increases the minimal scanning speed due to the increase in $\nabla T$ and the non-entropic force ${\bf f}_{s}$ (Eq. 69). It would be interesting to experimentally investigate a change of the direction of vortex motion as a function of the hot spot velocity and laser power (hot spot temperature). 

We would like to highlight that the non-entropic forces related to the temperature-dependent magnetization currents (in general, quantum currents) are completely different from the thermal forces, related to the entropy. In accordance with the Onsager approach, the all non-entropic forces created by $\nabla T$ are always equilibrated in quasi-equilibrium conditions and do not manifest themselves in the thermomagnetic phenomena. However,   some of none-entropic forces and corresponding phenomena can be observed in strongly nonequilibrium conditions, such as local or transient temperature gradient. In particular, vortex motion from the cold to hot area due to non-entropic force generated by mobile hot spot has been observed in Ref. \cite{44}.

\subsection{Thermomagnetic Effects in BKT State of 2D Superconductors}
Thermomagnetic phenomena in various 2D superconductors have attracted significant attention during last years. The Nernst coefficient has been measured in strongly disordered thin films of Nb$_{0.15}$Si$_{0.85}$,\cite{81} and amorphous films $\alpha$-MoSe,\cite{82} in flakes of crystalline NbSe$_2$,\cite{83} and in superconducting heterostructures, such as 2D Nb-doped strontium titanate (STO) placed between undoped STO cap and buffer layers\cite{82}  and in single-layer FeSe film on STO.\cite{84} It was found that above in 2D conductors the Nernst signal is observed in a wide temperature, which strongly exceeds the fluctuation region of traditional superconductors. In particular, the Nernst coefficient 2D NbSe$_2$ flakes was found to be at least two orders in magnitude larger than that in the bulk material.\cite{83}  All 2D superconductors show non-monotonic temperature dependence with the peak of the Nernst signal slightly below the transition temperature, where thermomagnetic effects are associated with vortices. There is no peculiarity of the Nernst coefficient at the transition temperature determined from the resistivity measurements.  Recent work\cite{82} has analyzed thermomagnetic characteristics related to 2D superconductors belonging to different families and found that the maximum Nernst signal corresponds to the entropy per vortex of 
\begin{equation}
{S}^*_{\phi} = k_B \cdot \ln 2.
\label{73}
\end{equation} 
As it is highlighted in Ref. \cite{82}, the entropy ${S}^*_{\phi}$ is approximately fifty times smaller than the intrinsic entropy of the vortex core with respect to superconducting background given by Eqs. 66 and 68. 

All results above show strong evidence in favor of the Berezinskii - Kosterlitz - Thouless (BKT) type superconductivity,\cite{85,86} which is well established in 2D superconducting heterostructures\cite{87,88} and thin disordered films, especially in materials with low superfluid density near the superconductor-insulator transition.\cite{89,90,91} In the BKT superconductor, a vortex created by magnetic field moves in the dense media of vortex - antivortex pairs and, therefore, the intrinsic entropy of the vortex with respect to the BKT background is negligible. This fact is well understood in the BKT theory, which takes into account only configuration entropy of dissociated vortex-antivortex pairs. Therefore, the transport vortex entropy, ${S}^*_{\phi}$, may be associated with the configuration entropy transfered by the vortex in the media of vortex-antivortex pairs. The factor of $\ln 2$ in Eq. 73, corresponds to the two positions available to the vortex in the real space, when the vortex and the pair exchange their positions. It is also intriguing, that according to Ref. \cite{82}, for multi-layer high-T$_c$ cuprates, such as La$_{1.92}$Sr$_{0.08}$CuO$_4$ studied in Ref. \cite{92}, Eq. 73 provides the thermomagnetic transport entropy per vortex per layer. This observation is well aligned with the pancake vortex structure in these materials. Our consideration predicts that the same thermomagnetic phenomena should be observed in the BKT state of Josephson junction arrays and studies of the transport entropy in corresponding experiments are very desirable. 

Our vorticity-based interpretation of thermomagnetic phenomena in thin films of disordered superconductors is well aligned with recent studies of the Nernst effect near the superconductor - insulator transition (SIT).\cite{93,94} Due to vorticity a large Nernst signal is observed on both the superconducting and the insulating sides with a signal peak close to the critical point. However, the nature of the transport entropy transfered by vortices near SIT is still under investigations.\cite{94} 

\section{Electrical Readout with Perfect Thermal Isolation for Quantum Sensors}

Quantum currents, - magnetization, superconducting, persistent and topological edge currents, - transfer the electric charge and spin, but not entropy. This unique combination of thermal and electric/spin properties is valuable for many applications related to quantum computing, networking, and sensing. 

Resistive and kinetic-inductance micro and nanobolometers are used as direct detectors, wide band mixers, single photon counters, and quantum colorimeters (for a review see Ref. \cite{95}). Andreev mirrors formed at the interface of a micro/nano-sensor and superconducting interconnectors are widely used to prevent cooling of  ultra-sensitive THz and IR hot-electron nanobolometers.\cite{96,97} To achieve high sensitivity, the detector should be very small and exceptionally well thermally isolated from the environment, including the heat transfer to connectors related to the electric readout. Fast out-diffusion of hot electrons from the nanoscale sensor leads to fast sensor cooling, which reduces the detector sensitivity. Andreev mirrors 
eliminates cooling of hot electrons in the sensor via interconnectors. Thermal processes at the interface between a normal metal and a superconductor have been studied by A. F. Andreev to describe the anomalous thermal resistance of a superconductor in the intermediate state.\cite{98} Electron scattering from the interface involves conversions between electron and hole Fermi liquid excitations to a Cooper pair, which forms a superconducting condensate. As superconducting condensate does not transfer entropy and. therefore, superconducting nanowires are ideal interconnectors that provides electric coupling  without thermal coupling. 

Inter-connectors based on Quantum Hall edge states in topological insulators may be employed to semiconductor sensors such as the hot electron bolometer based on radiation-induced intersubband transitions in a high mobility quantum wells at nitrogen temperatures.\cite{99} The sensitivity of this detector is determined by the cooling time of hot electrons excited by the radiation from the first subband and redistributed between between subbands due to inter-electron interaction. Due to high mobility of electrons ($\sim 5\cdot 10^3$ cm$^2$/ V s) the out-diffusion cooling becomes important at 10 $\mu$m length of the quantum well detector. Thus, the micron and sub-micron length detectors require quantum mirrors such as quantum Hall edge states to prevent the out-diffusion cooling. 
  
Polarization-sensitive detectors have a potential to enhance sensing technologies. Such detectors can employ quantum wells with spin polarized states in subbands, which have different electron mobilities.  So, the electron transitions between spin-polarized states due to polarized radiation change the device resistance, which is measured by a readout. To prevent out-diffusion cooling of spin polarized detectors one can use quantum spin Hall helical edge states protected by Kramers’ theorem.\cite{100} The quantum phase coherence of wave functions in spin Hall helical edge states prevents the heat/entropy transfer by these spin current inter-connectors. 

\section{Discussion}

In this work we have investigated interplay of transport thermomagnetic currents and quantum currents in charge, energy, and heat (entropy) fluxes.  In a small electric field or temperature gradient, quantum currents do not generate and transfer intrinsic entropy.   For example, magnetization energy flux, $c {\bf M} \times {\bf E} $, does not contribute to the heat (entropy) current. In Section II we support this rather obvious statement with a textbook example of the thermomagnetic coefficient of noninteracting electrons in quantized magnetic field. Even in this rather simple case, the ``magnetization heat current'' $c {\bf M} \times {\bf E} $ would lead to the giant Nernst coefficient. Thus, the transport heat current in crossed electric and magnetic fields equals the bulk heat current without any magnetization corrections. 

The bulk electric current has the magnetization component, $-c (d \bf{M} /dT) \times \nabla T$.  According to Obraztsov\cite{54}, this component should be removed from the the bulk electric current to obtained the transport current (Eq. 26). As Onsager relation is valid for electric and heat transport currents, the Onsager relation for the bulk thermomagnetic currents is not symmetric with respect to magnetization. Alternatively, the transport electric current in crossed magnetic field and temperature gradient may be calculated as a response to entropic thermal forces induced by the temperature gradient (Eq. 17). As the temperature-dependent magnetization does not generate entropy and does not produce entropic thermal forces, the magnetization contribution is absent here. Approach based on entropic forces is widely used for the vortex thermomagnetic transport.\cite{49} In this case, entropy flux and entropic forces directly provides heat and electric transport currents.  

Nonlocality of magnetization currents is well understood.\cite{42} Magnetization currents are circulating and divergent free and, therefore, the net electric magnetization current through any conductor cross-section is zero. Cancellation of bulk and surface magnetization contributions in the net charge and heat flows is a direct consequence of the Maxwell equations. It is applicable to any geometry of the conductor including the Corbino disk geometry. Thus, the net electric and heat thermomagnetic currents equal transport electric and heat currents and satisfy the same Onsager relations as transport currents.   

Let us highlight, that zero entropy of magnetization currents is a direct consequence of basic classical and quantum concepts. In thermodynamics, the entropy is calculated as a temperature derivative of the free energy with respect to temperature at constant magnetization (Eq. 9) and, therefore, the entropy does not consists of terms proportional to $d \bf{M} /dT $, i.e. to the magnetization current, $-c (d \bf{M} /dT) \times \nabla T$. 
Also, the surface magnetization currents, $ c {\bf M}\times {\bf n}$ (${\bf n}$ is the unit vector normal to the surface),  present the macroscopic motion of a group of electrons and, according to thermodynamics (e.g. see \S 10 of Ref. \cite{23}), the kinetic energy and electromagnetic energy related to this motion do not contribute to the internal energy, free energy, and entropy. In particular, in the Corbino disk geometry the magnetic filed perpendicular to the disk plane and the radial electric field generate the azimuth bulk and edge magnetization electric currents. In the rotating frame, these currents are absent and corresponding entropy is absent. As entropy is an invariant with respect to rotation with constant angular
velocity,\cite{23} the magnetization currents have zero entropy in the usual frame. 

Quantum thermodynamics also gives zero entropy of magnetization currents (quantum currents). In the resource theory of coherence\cite{101}, zero entropy is a direct consequence of the rigidness of the wave function and may be derived in a way similar to London's approach to the wave function in a superconductor. In the dynamical view of quantum thermodynamics based on the theory of open quantum systems,\cite{102} the thermal forces affected electrons in thin films may be introduced via the electron interaction with substrate phonons, which are placed in the temperature gradient (the short phonon mean free path is required to define a smooth temperature gradient.\cite{42,52}) As electrons in quantum currents are not scattered by phonons, any drag effects of quantum currents by phonons are absent and the entropic thermal forces are also absent. Thus, both classical and quantum concepts lead to zero magnetization entropy,  zero magnetization heat flux in crossed electric and magnetic fields, and zero thermal forces related to magnetization currents in crossed temperature gradient and magnetic field. This  description satisfies the Onsager relation. 

While the net magnetization energy current is always zero, its bulk and surface counterparts in magnetic materials and metamaterials can be very significant. To calculate the Poynting vector, the bulk magnetization energy flux, $c {\bf M} \times {\bf E} $, should be added to the electromagnetic energy flux, $c [{\bf E} \times {\bf H}]/4 \pi $. Thus, the Poynting vector is given by $c [{\bf E} \times ({\bf H }  - 4\pi{\bf  M})]/ 4 \pi$ (Eq. 33).  We have shown that the term ${\bf j}^e_{mag} \cdot {\bf E}$ previously associated with the dissipated power (see Refs. \cite{15,39,40}) in fact, describes the energy redistribution between the surface and bulk energy currents (see Fig. 3). The new expression for the Poynting vector and dissipationless redistribution of the magnetization energy are critically important for understanding of the energy transfer in metamaterials. 

Besides magnetization currents, we also investigate currents created by fluctuating Cooper pairs above the transition. In strong magnetic fields, $\omega_c \tau \gg 1$, the thermomagnetic transport is collisionless and thermomagnetic coefficient is directly presented via the entropy of electron system (Eq. 12). We obtained thermomagnetic coefficient of fluctuating Cooper pairs for various relation between temperature and magnetic field (Eq. 39). As all fluctuation corrections to the electron entropy are always small (Eq. 38), the corresponding corrections to the thermomagnetic coefficient are also small. 

In general case with respect to the parameter $\omega_c \tau$, many-body corrections to the thermomagnetic coefficient may be calculated by the Kubo method. As magnetization currents do not contribute to the heat current, the heat current - electric current correlator does not contain any magnetization component and, therefore, the Kubo approach does not require any magnetization correction. In Section VII, using the gauge invariant formalism developed for the Hall coefficient in the fluctuating region,\cite{43} we have calculated the thermomagnetic coefficient at small magnetic fields and show that it scales as $\omega_c \tau \cdot T/\epsilon_F$ (Eq. 57), i.e. as the normal state thermomagnetic coefficient that combines the Hall angle $\omega_c \tau$ and the Fermi liquid $T/\epsilon_F$ factor in the electron entropy. 

Let us note that our microscopic calculations are based on entropic approach and corresponding ``$\epsilon$ -presentation'' (``$\omega$ -presentation'') of the heat/entropy flux, where $\epsilon$ ($\omega$) is an argument in the fermion (boson)  distribution function. In this approach the thermal energy equals $\epsilon$  ($\hbar \omega$), because in the linear approximation the functional derivative of the entropy with respect to the distribution function is $\epsilon/k_B T$  ($\hbar \omega/k_B T$).\cite{13} The entropic approach directly defines the thermal energy via statistics and does not require any corrections related to magnetization energy and significantly simplify calculations. As it was highlighted by Johnson and Girvin,\cite{106}, by neglecting the magnetization energy fluxes in the energy balance, one would obtain huge thermoelectric coefficient in quantized magnetic field with the characteristic ``thermal energy'' of the order of the magnetic energy $\hbar \omega_c$. 

In Section VIII we develop the entropy - based theory of the thermomagnetic transport in the Fermi liquid at arbitrary values of $\omega_c \tau$. In the approximation of energy-independent electron mobility we derive exact thermodynamic equation for the thermomagnetic coefficient (Eq. 59), which corrects empiric formulas for the Nernst coefficient obtained in recent works \cite{34,35} and \cite{36}.
We also introduce the thermomagnetic entropy per unit charge as the entropy per carrier averaged over the product of longitudinal and transverse mobilities (Eq. 63) and show that the Nernst coefficient is proportional to the difference between thermoelectric and thermomagnetic entropies per charge (Eq. 65). If the electron mobility (momentum relaxation time) is isotropic and energy independent, both thermoelectric and thermomagnetic entropies per charge are the same (Eqs. 61 and 63) and Eq. 65 gives the Sondheimer cancellation in terms of the transport entropy. In the case of  $\omega_c \tau \gg 1$ the Nernst coefficient is also negligible, because the thermoelectric entropy per charge equals the thermomagnetic entropy.  Specific many-body corrections to the entropy and mobility of Fermi liquid quasiparticles may be incorporated in this approach. 

In accordance with thermodynamics of irreversible processes, in the linear in electric field (temperature gradient) approximation all non-entropic forces are always canceled among themselves and, as we discussed above, contributions of quantum currents to the net electric current are absent (see Section III). However, in strongly nonequilibrium conditions, the non-entropic forces can be unbalanced. In particular, the non-entropic thermal force that affect superconducting electrons placed in the transient temperature gradient leads to the vortex motion from the cold to hot area. In Section IX we explain results of recent experiments with single vortices, which follow to the hot spot created by laser beam.\cite{44} We also analyze thermoelectric transport in the BKT state of 2D superconductors, such as thin disordered films and superconducting heterostructures, and show that the transport entropy transfered by a single vortex over the background of vortex - antivortex pairs is given by the configuration entropy, $k_B \cdot \ln 2$ (Eq. 73).  This result explains experimental data of the recent work\cite{82} and other previous publications. 

Quantum currents transfer the electric charge and spin, but not entropy. This unique combination of thermal and electric properties of superconducting currents is already used in many applications. Quantum Hall edge currents and quantum spin Hall edge currents also do not transfer entropy and have a strong potential to enhance sensing applications by preventing electron cooling in nanosensors. 

\section*{Acknowledgements}
We thank I. Aleiner, P. Ong, Y. Kopelevich, K. Behnia, N. Noginova, and B. Karasik for valuable discussions.

\end{document}